\documentclass{revtex4}
\begin{document}

\title{Measurement of the total energy of an isolated system by an
internal observer}
\author{S. Massar}
\affiliation{Laboratoire d'Information Quantique and QUIC, {C.P.} 165/59, 
Universit\'{e}
Libre de Bruxelles, Av. F. D. Roosevelt, B-1050 Bruxelles,
Belgium}  
\author{S. Popescu}
\affiliation{H. H. Wills Physics Laboratory, University of Bristol, Tyndall 
      Avenue, 
      Bristol BS8 1TL, U.K.}
\affiliation{Hewlett-Packard Laboratories, Stoke Gifford, Bristol BS12
  6QZ, U.K.} 
\date{\today}
\begin{abstract}
We consider the situation in which an observer internal to an isolated
system wants to measure the total energy of the isolated system
(this includes his own energy, that of the measuring device and clocks
used, etc...). We
show that he can do this in an arbitrarily short time, as measured by
his own clock. This measurement is not subjected to a time-energy
uncertainty relation. The properties of such measurements
are discussed in detail with  particular emphasis on the 
relation between the duration of the
measurement as measured by internal clocks versus external clocks.
\end{abstract}

\maketitle

\section{Introduction}

The interpretation of quantum mechanics is based to a large extent
on understanding the measurement process. This is because
measurements are the interface between quantum systems and the
observer. The present paper is devoted to the study of
measurements in a very particular situation, namely we shall
consider an observer inside an isolated system who attempts to
measure the total energy of the isolated system. We shall be
particularly interested in the time it takes the internal observer
to measure the total energy. Our analysis sheds light on
several troublesome aspects of quantum mechanics: the nature of
observable quantities in quantum mechanics, the interpretation of
energy measurements, the interpretation of
isolated systems and the interpretation of time for such systems. 

A central theme in our paper is the interpretation of the time-energy 
uncertainty relation in the context of energy measurements. Before delving
into the specific subject of this article it is useful to recall what
is known about this question. It was initially believed (in analogy
with the situation for position and momentum analyzed by Heisenberg)
that the precision $\Delta E$ with which the energy of a quantum
system is measured and the time $T$ taken by the measurement
are related by $T \Delta E  \geq 1$ (where we take $\hbar =1$)
\cite{LP}. However it was shown by Aharonov and Bohm \cite{AB} that
this intuition is incorrect and that the energy of a quantum system
can be measured in arbitrarily short time. (For a collection of
reprints on quantum measurements, including the above two articles,
see \cite{WZ}). Recently it was realized
that in order to carry out the fast measurement of Aharonov and Bohm
the Hamiltonian must be known before hand. If the Hamiltonian is not
known then the energy cannot be determined instantaneously, and in
this case the precision $\Delta E$ with which the energy of a system
is measured and the time taken for this measurement are constrained
by $T \Delta E \geq 1$\cite{AMP}.

In the works described above the measurement is carried out by an
external observer who couples the system to be measured to an external
measuring device for a time $T$. Aharonov and Reznik \cite{AR}
added a twist to this problem by asking whether an observer internal
to an isolated quantum system can measure the total energy of the
system itself. They showed that such internal measurements of the
total energy are indeed possible. Aharonov and Reznik illustrate this 
by a simple example in which a self gravitating system, say a planet
of radius $R$, ejects outwards a small mass $m$. By measuring the time
$t$ it takes the mass to fall back onto the planet one can
determine the mass of the planet and thus 
deduce its total energy $E$ 
through the correspondence between energy and mass $E= M c^2$.
Aharonov and Reznik also present a more abstract mathematical model
to illustrate this, see section \ref{II} for a detailed presentation
of this model.

One of the main points of the present work is to show that in
discussing the time taken to measure the energy of an isolated system
one should make a distinction between the time measured by the
external observer (the external time $t_{ext}$) and the time measured
by the observer internal to the system (the internal time
$t_{int}$). In particular when discussing the time energy uncertainty
one should distinguish whether the duration of the measurement 
$T$ is measured in internal or external time. Note that
having the time of the internal and external observer differ is
completely compatible with the usual principles of physics, and in
particular with the general relativity where different observers have
in general completely different time variables. We discuss this point
further below.

The works \cite{AB,AMP,AR} complemented with the results obtained in
the present article show that the relationship between the precision
$\Delta E$ with which the energy of a system is measured and the time
$T$ taken for this measurement is much richer than previously
thought. Indeed there are many cases which can be considered according
to whether the observer which carries out the measurement is internal
or external to the  system, according to whether the duration of the
measurement is measured in internal or external time, and according to
whether the  Hamiltonian of the system is known or unknown. Table
\ref{T1} summarizes these different possibilities as well as what is
known about them. 
Case 1 in the table is the situation analyzed in \cite{AB}, case 2 is
the situation analyzed in \cite{AMP} and case 7 is the subject of the
present article. We will also make some comments about case 5. 
Note that little or nothing is known about most of
the cases in the table. Thus much more work is required to completely
understand the status of the time-energy uncertainty relation in the
context of energy measurements.

\begin{table}\label{T1}
\begin{tabular}{c|c|c|c|c|}
\hline
& Observer who carries
 & Time variable for which & Hamiltonian of the system & Constraint on
duration   \\
& out the measurement & the duration of the & (known or unknown)& of
the measurement \\
&(external or internal)& measurement is minimized & & \\
& & (external or internal time) & & \\
\hline
1& external & external & known & $T_{ext}$ arbitrarily small \\
\hline
2& external & external & unknown & $T_{ext} \Delta E \geq 1 $ \\
\hline
3& external & internal & known & ? \\
\hline
4& external & internal & unknown & ? \\
\hline
5& internal & external & known & ? \\
\hline
6& internal & external & unknown & ? \\
\hline
7& internal & internal & known & $T_{int}$ arbitrarily small \\
\hline
8& internal & internal & unknown & ? \\
\hline
\end{tabular}
\caption{Summary of relations between the precision $\Delta E$ with
  which the energy of a system is measured and the time $T$
  taken for the measurement, according to whether 
the observer which carries out the measurement is internal
or external to the  system, according to whether the duration of the
measurement is measured in internal or external time, and according to
whether the  Hamiltonian of the system is known or unknown.}
\end{table}

Let us now turn back to the case of the internal observer who wants to
measure the total energy of an isolated system and who knows the
Hamiltonian of the isolated system. In the particular examples they analyze
Aharonov and Reznik find an intriguing effect, namely 
that the precision $\Delta E$ with which the
total energy is measured and the time taken for the measurement are
not independent. They find that they are related by $T \Delta E 
\geq 1$. On the basis of these examples they go on to argue that this
is  a fundamental constraint relating the precision with which the
total energy of an isolated system is measured and the amount of time
taken to carry out the measurement.  A central result of the present
article is to correct this statement.

Let us first note that if Aharonov and Reznik's claim was to be
confirmed it would have important consequences for the interpretation
of observable quantities in quantum mechanics because it would mean
that there are quantities which are observable in principle, but are
not observable instantaneously because they can only be measured in a
finite time. This could give rise to some surprising (even paradoxical)
situations \cite{A} and is contrary to the situation concerning all other
measurements of observables. For instance it is quite obvious that a
position measurement can be done arbitrarily fast. And it was
shown in \cite{AB} that any observable, and in particular the
Hamiltonian observable, can be measured in arbitrary short time.
(Note that the instantaneous measurements described in\cite{AB}
require a measuring device with arbitrarily large energy external to
the system to be measured. They therefore cannot be 
applied to measurements of the energy of an isolated system by an
internal observer).

A clear understanding of measurements of total energy by an
observer inside an isolated system therefore have direct bearing
on the interpretation of quantum mechanics and in particular on
what is meant by an observable quantity. In this paper we shall
resolve the apparent paradox discovered by Aharonov and Reznik. To
this end we must, as mentioned above, 
distinguish two notions of time for isolated
systems. First there is the internal time $t_{int}$. This is the time as
measured by a clock inside the isolated system. We shall show by
an example that the precision $\Delta E$ with which the internal
observer measures the total energy of the isolated system and the
amount of internal time taken for the measurement $T_{int}$
are not mutually constrained. In particular $T_{int}$ can
be much smaller than $1/\Delta E$, see case 7 in the table. 
(Throughout this article we will denote by the capital letter T the duration of
the measurement and by the small letter t the time variable.)

Second there is the external time $t_{ext}$. This is the time which
would be measured by an observer outside the system. We do not know
whether the precision $\Delta E$ with which the internal
observer measures the total energy of the isolated system and the
amount of external time taken for the measurement $T_{ext}$
are mutually constrained, although it is tempting to conjecture, see
section \ref{IV}, that in this case $T_{ext} \Delta E \geq 1$
holds. At present however we leave a ? in case 5 in the table.

The resolution of the apparent paradox uncovered in \cite{AR} is
therefore that the internal observer can measure the total energy
arbitrarily fast in his own proper time. Similarly the total
energy of the isolated system can be measured by the external
observer arbitrarily fast \cite{AB}. Thus {\em the energy of an isolated system
can be measured arbitrarily fast in the proper time of the
observer, whether or not he is internal to the system or not.}

As mentioned above having the internal and external times differ
isn't in contradiction with basic physical principles. 
Indeed in general relativity the
clocks of two systems have no reason to be synchronized, or to stay so
during their evolution. We can thus imagine that the isolated system
is a planet. For an observer at
the surface of the planet and an observer at infinity, the times
differ by the gravitational red-shift factor: $t_{surface}=
\sqrt{g_{00}}t_\infty$. If $g_{00}$ is constant this is just a
rescaling of time. But if $g_{00}$ changes during the energy
measurement itself, than the amount of external time $T_\infty$ and
the amount of internal time $T_{surface}$ 
taken for the measurement may differ in a non trivial way. Thus we can
imagine that 
the measurement begins by a dynamical evolution during
which $g_{00}$ decreases from its initial value to a very small
value. This could for instance be due to a gravitational collapse
of the planet which stops just above its Schwarzschild radius. 
 Once $g_{00}$ is small the
internal observable carries out the measurement of his total
energy. Once the measurement is finished the planet expands and
brings $g_{00}$ back to its initial value.
This shows that the measurement can have very different durations for
the internal and external observer. In section \ref{III} we shall show
that the duration for the internal observer can be arbitrarily short.

Furthermore $g_{00}$ is in fact an operator and can have uncertain
value. This means that the relation between $t_{surface}$ and
$t_{\infty}$, and hence the relation between internal and external
time, becomes uncertain. This effect is inevitable and plays an
essential role in understanding these measurements, as we discuss below.

In what follows we first review the mathematical model of Aharonov and
Reznik and point out its limitations. We then introduce an
alternative model in which the energy of an isolated system can be
measured 
by an internal
observer arbitrarily fast in his internal time. Finally we discuss its
interpretation. 

\section{The model of Aharonov and Reznik}\label{II}

Aharonov and Reznik consider an isolated system containing a clock.
The system is described (in the terminology of \cite{AR}) by the
Hamiltonian $H_c + H_{box}$ where $H_c= -i \partial_x = p_x$ is the
Hamiltonian of the internal clock variable $x$ ($[x, p_x]=i$) and
$H_{box}$ is the  
Hamiltonian of the rest of the isolated system. For simplicity we take
$H_{box}$ to be independent of $x$, but the analysis can be easily
generalized to the case where it depends on $x$. 

In addition the isolated system contains a measurement variable $q$ with
conjugate momentum $p$. The outcome of the measurement will be
recorded in the value of $p$. The measurement only takes place for $x_i\leq
x \leq x_f$. Thus for $x<x_i$ or $x>x_f$ the measuring device is
uncoupled to the system and the total Hamiltonian is $H_c + H_{box}$
(we take for simplicity the free Hamiltonian of the measuring device
to be zero) whereas between $x_i \leq x \leq x_f$ the measuring
device is coupled to the system. Aharonov and Reznik take the coupling
to be a von-Neumann type interaction\cite{vN,Bohm} of the
form\footnote{Recall that for measuring a variable $A$ the standard
  von Neumann interaction Hamiltonian is $g(t) A q$ where $t$ is the
  time variable, ie. a classical parameter. In our case however the
  role of time is played by the clock pointer $x$, so that $g(t)\to
  g(x)$. Furthermore in our case $A= H_c + H_{box}$. Since $g(x)$ does
  not commute with $H_c$ the interaction Hamiltonian has to be
  symmetrised: $g(x) H_c \to (g(x) H_c + H_c g(x) ) /2$.}:
\begin{equation}
H = H_c + H_{box} + \left(
\frac{1}{2} g(x) H_c + \frac{1}{2} H_c g(x) + g(x) H_{box}\right) q
\ .\label{HAR}
\end{equation}
Here $g(x)$ is the coupling function which is non zero only when $x_i
\leq x \leq x_f$. We take $g(x)$ to increase rapidly from zero to $g$
when the measurement starts, then to stay constant during the
measurement, and to decrease rapidly to zero when the measurement
ends, see fig. 1.  The duration of the measurement, as measured by the
$x$ variable, will be denoted $L= x_f - x_i$.

Let us suppose the system is in an eigenstate of the total energy
\begin{equation}
H|\Psi \rangle = E_0 |\Psi\rangle\ ,
\label{HPsi}
\end{equation}
that the box is in an eigenstate of $H_{box}$
\begin{equation}
H_{box}|u_{E_{box}} \rangle = E_{box} |u_{E_{box}}\rangle\ ,
\label{Hubox}
\end{equation}
and that the measuring device is in an eigenstate of the measurement
variable $q$. Then the state of the isolated system can be written as
\begin{equation}
|\Psi\rangle 
= \psi(x,E_0,E_{box},q) |q\rangle |u_{E_{box}}\rangle \ .
\label{Psi}
\end{equation}
Substituting into eq. (\ref{HAR}) one obtains the exact solution:
\begin{equation}
|\Psi\rangle 
= \frac{1}{\sqrt{1 + g(x) q}}e^{-iE_{box} x} e^{i E_0 \int^x
  \frac{dx'}{1 + g(x') q}} |q\rangle |u_{E_{box}}\rangle \ .
\label{PsiAR}
\end{equation}

The internal time is $$t_{int}=x$$ since this is the variable that
multiplies $E_{box}$. Indeed if one realizes a superposition of
different states with different $E_{box}$, then the internal state of
the box will evolve in time $x=t_{int}$.

On the other hand the external time, expressed as function of the
variables of the isolate system, is
$$t_{ext} (x,q) =  \int^x
  \frac{dx'}{1 + g(x') q}$$
since this is the variable that multiplies the total energy
  $E_0$. Indeed if one realizes a superposition of states with
  different values of $E_0$, then one will find that
  $t_{ext}(x,q)\simeq t$ (where
  the wavefunction is solution
  of $i\partial_t \Psi = H \Psi$, ie. $t$ is the absolute external
  time which can be measured to arbitrary high precision by an
  external observer with sufficient energy).

In order to analyze the solution eq. (\ref{PsiAR})
Aharonov and Reznik suppose that $g(x) q << 1$ and then expand the
phase in eq. (\ref{PsiAR}) to first order in $g(x)q$.
Note that when $g(x) q <<1$,
the external time and the internal time coincide $t_{ext}=t_{int}$ and
the two need no longer be distinguished. As we shall see in section
\ref{III} these times need not necessarily coincide. 

One of the points of the following discussion is to show that the
first order expansion of Aharonov and Reznik is not always
legitimate. To illustrate this we shall keep the second order terms in
the phase. Thus we approximate eq. (\ref{PsiAR}) as
\begin{equation}
|\Psi\rangle 
= \frac{1}{\sqrt{1 + g(x) q}}e^{-iE_{box} x} e^{i E_0 \int^x
  dx' (1 - g(x') q + g^2(x') q^2) } |q\rangle |u_{E_{box}}\rangle \ .
\label{PsiAR2}
\end{equation}
After the measurement , ie. when $x >  x_f$, and using the fact that
$\int_{x_i}^{x_f} dx' g^n(x') \simeq L g^n$ (see the form of $g(x)$ in
fig. 1) we have
\begin{equation}
|\Psi\rangle 
= e^{-iE_{box} x} e^{i E_0 x} e^{-i E_0 L g  q + i E_0 L g^2 q^2} 
|q\rangle |u_{E_{box}}\rangle \ , \ x > x_f \ .
\label{PsiAR3}
\end{equation}
Let us now suppose that initially the measuring device is in a
Gaussian state $N e^{-q^2/ 2 \sigma^2}$ centered on $q=0$ with width
$\Delta q = \sigma$ ($N$ is the normalization constant). Then at late
times the state becomes
\begin{eqnarray}
|\Psi\rangle 
&=& \int dq N e^{-q^2/ 2 \sigma^2} e^{i (E_0 -E_{box}) x} 
 e^{-i E_0 L g  q + i E_0 L g^2 q^2} 
|q\rangle |u_{E_{box}}\rangle \nonumber\\
&=&  N'  e^{i (E_0 -E_{box}) x}
\int dp \exp\left[ - \frac{(p - Lg E_0)^2}{2} \frac{\sigma^2}{1 + 4 L^2 g^4
E_0^2 \sigma^4}(1 + i 2 L g^2 E_0 \sigma^2)\right]
|p\rangle |u_{E_{box}}\rangle
\label{PsiAR4}
\end{eqnarray}
where we have rewritten the state in momentum representation.

Thus we see that the mean value of $p$ has been displaced by 
\begin{equation}\label{A}
p \to p + L g E_0\ .
\end{equation}
This means that a measurement of the total energy of the isolated
system has indeed taken
place and that its value is registered in the pointer $p$.

The spread in the value of $p$  is given by eq.(\ref{PsiAR4}):
$$
\Delta p = \frac{1}{\sigma}\sqrt{1 + 4 L^2 g^4 E_0^2 \sigma^4}\ .
$$
The precision $\Delta E_0$ with which the energy is measure is thus
\begin{equation}\label{precAR}
\Delta E_0 =  \frac{1}{L g\sigma}\sqrt{1 + 4 L^2 g^4 E_0^2 \sigma^4}\ .
\end{equation}

In discussing  the precision of the measurement we must thus
distinguish two cases according to the values of $E_0$.
\begin{enumerate}
\item
$|E_0| < 1/ 2 L g^2 \sigma^2$. In this case $\Delta E_0 \simeq 1/ L g
\sigma$. We recall that $g\sigma << 1$ (this is the condition for the
expansion in $g(x) q$ in the phase to be valid) and that $L
=T$ is the duration of the measurement (we do not distinguish
between internal and external time since when $g \sigma << 1$ they
coincide) to obtain
\begin{equation}\label{ineqAR}
T \Delta E_0  >> 1 \ .
\end{equation}
Note that one can come close to saturating this relation when $g
\sigma$ is not very small compared to $1$.
\\
\item $|E_0| > 1/ 2 L g^2 \sigma^2$.
In this case it is the second term under the square root in 
eq. (\ref{precAR}) which
dominates and one cannot even come close to saturating
eq. (\ref{ineqAR}). In fact when $|E_0|$ increases one is further and
further from saturating eq. (\ref{ineqAR}). 
\end{enumerate}

Several remarks are now in order. First of all the parameters $g,
\sigma, L$ can be chosen by the internal observer. But once they are
chosen they are fixed. This means that there will be only a finite
band of energy for which eq. (\ref{ineqAR}) can be approximately saturated.
As the energies get further and further from this band one is further
and further from saturating this inequality.
This means that 
eq. (\ref{ineqAR}) does not have a fundamental character in the
context of the model of \cite{AR}.

Second the initial state chosen above $e^{-q^2/2 \sigma^2}$ implies
that the resolution of the measurement is maximal near $E_0=0$. By
putting a phase on the initial state one can change the energy at
which the which the resolution comes close to saturating
eq. (\ref{ineqAR}). But it will always be optimal in an energy band
only.

Third in the above calculation we took a Gaussian initial state and
expanded the phase to second order so as to be able to do an exact
calculation. It is not difficult to persuade oneself -for instance by
using an analogy with the effect of dispersion on propagating waves-
that 
the general conclusion, namely that the time
energy uncertainty relation eq. (\ref{ineqAR}) can only be saturated
in an energy band, continues to hold for arbitrary states and
when the phase is expanded to higher orders in $gq$. 

In summary there are two problems with the conclusions of \cite{AR}
concerning the time energy uncertainty.  First of all it is only a 
model, and one
should not draw universal conclusions about the validity of the
time-energy uncertainty from a single model. Secondly in this model
the time-energy uncertainty, although it is obeyed, does not appear as
fundamental since for most values of the energy is cannot be saturated.

\section{A model which allows internal observers to measure the total
  energy arbitrarily fast }\label{III}

We now introduce an alternative model that allows an internal observer
to carry out 
measurements of the total energy arbitrarily fast in his own internal
time. 
The Hamiltonian of the box, the clock and the measuring device is:
\begin{equation}
H = \frac{H_{box}}{1 + q g(x)} + 
\frac{1}{2} \left( \frac{1}{1 + q g(x)} H_c + H_c\frac{1}{1+q  g(x)} \right) 
\label{HMP}
\end{equation}
where the notation is the same as in section \ref{II}, eq. (\ref{HAR}).

Note that this Hamiltonian can be thought of as describing the
isolated system undergoing gravitational collapse as discussed in the
introduction. Indeed the Hamiltonian of such a collapsing planet would
be $H=\sqrt{g_{00}}H_{planet}$. Our model thus corresponds to taking
the metric to be $\sqrt{g_{00}}= 1/(1 + q g(x))$. Note that this
metric is an operator (since $q$ and $x$ are operators). This will
play a crucial role in what follows.

The exact solution of the Schr{\"{o}}dinger equation $H|\Psi \rangle = E_0
|\Psi\rangle$ is
\begin{equation}
|\Psi\rangle 
= \sqrt{1 + g(x) q}e^{-iE_{box} x} e^{i E_0 \int^x dx' (1 + g(x') q)} 
|q\rangle |u_{E_{box}}\rangle 
\label{PsiMP}
\end{equation}
where the notation is once more as in section \ref{II}, eq. (\ref{PsiAR}).

In the present case the internal time is 
\begin{equation}
t_{int} = x \label{tintx}
\end{equation}
 and the
external time, expressed as function of the
variables of the isolate system, is
\begin{equation}
t_{ext}=\int^x dx' (1 + g(x') q)
\label{textx}
\end{equation}
since these are the functions that multiply $E_{box}$ and $E_0$
respectively in the phase of $\Psi$. 
Note that before and after the measurement $g=0$ and the internal and
external time coincide. But we will allow them to be very different
during the measurement.

After the measurement, when $x > x_f$, the solution takes the form
(where we use the form of $g(x)$ given in fig. 1):
\begin{equation}
|\Psi\rangle 
= e^{-iE_{box} x} e^{i E_0 x} e^{i E_0 L g  q }
|q\rangle |u_{E_{box}}\rangle \ , \ x > x_f \ .
\label{PsiMP3}
\end{equation}
Thus after the measurement the momentum $p$ of the measuring device
has been displaced by
\begin{equation}\label{AMP}
p \to p - L g E_0\ .
\end{equation}
Note that no approximations have been made in obtaining this result,
contrary to the the way
eq. (\ref{A}) is obtained in the model of Aharonov and Reznik.
If the momentum of the measuring device has spread $\Delta p$, then
eq. (\ref{AMP}) constitutes a measurement of total energy with
precision 
\begin{equation}
\Delta E_0 = \frac{\Delta p}{Lg}
\ .
\label{DEMP}
\end{equation}

In order to interpret eq. (\ref{DEMP})
we must specify  the initial state of the measuring device. We will
take the initial state to only have values $q > 0$. This ensures (since $g>0$) 
that the
Hamiltonian eq. (\ref{HMP}) never becomes infinite and that the model
is well defined. We will thus suppose that the initial state
is
centered on $\overline q$
with width $\Delta q$
such that $\overline q >> \Delta q$.
Note that these constraints are compatible with taking 
the initial state to be almost Gaussian and thus
to almost saturate the inequality
 $\Delta q\Delta p \geq
1$. 
The constraint on the 
precision of the measurement can be rewritten as 
\begin{equation}
\Delta E_0 \geq  1/ (  L g
\Delta q )
\label{DEMP2}\end{equation}  and this inequality can be almost
saturated in the case of the quasi Gaussian states just mentioned.

The amount of internal time it takes to carry out the
measurement is $T_{int} = L$. We can choose $L$ arbitrarily
small. We can also choose $g$ such that $L g \Delta q$
is arbitrarily large. Thus the internal observer can measure the total
energy of the system to arbitrarily high precision in arbitrarily
small internal time. This is our main result.

\section{The relation between internal and external time.}\label{IV}

Now we come to an interesting point, namely how the external time
variable $t_{ext}$ is related to the internal time variable $t_{int}$
and how the duration of the measurement, as measured by the two
observers, are related.
Let us imagine that before the measurement starts both clocks are
synchronized and that the measurement starts at $t=0$ as indicated by
both clocks. As soon as the measurement
is finished, the internal observer sends a signal to the external
observer. What will be the time $T_{ext}$ indicated by the external clock?

To answer this question we use eqs. (\ref{tintx}) and (\ref{textx}) to
find
\begin{equation}
T_{ext} = T_{int} + T_{int} g q\ .
\label{TT}
\end{equation}
$T_{int}$ is well defined. However
$q$ is an operator and has uncertainty. Therefore $T_{ext}$ is uncertain.
This raises two questions: what is its average value
$\overline{T_{ext}}$ and
what is its spread $\Delta T_{ext}$? 

The uncertainty is given by
$\Delta T_{ext} = L g \Delta q$. The constraint
eq. (\ref{DEMP2}) can thus be interpreted as a relation between the
precision of the measurement and the uncertainty in the duration of
the measurement as
measured in external time:
\begin{equation}
\Delta E_0 \Delta T_{ext} \geq 1
\ .\label{DEDTMP}\end{equation}
We will argue below that this constraint is universal
and 
must apply to
all measurements of total energy by an internal observer.

The average duration of the measurement is $\overline{T_{ext}}=
T_{int} (1+ g \overline{q})$. 
Since in our model $\Delta q << \overline q$ the average duration
of the measurement, as measured in external time, is constrained by
\begin{equation}
\Delta E_0 T_{ext} \geq 1 .
\label{DETMP}
\end{equation}
However as discussed below we are not sure whether this constraint is
universal.

We now address the universality of eqs. (\ref{DEDTMP}) and (\ref{DETMP}).
To this end let us view the {\em total} system consisting of
the internal {\em and} the external system as a whole. Let us suppose
that the total system is in an energy eigenstate. Let us also suppose
that initially the internal and external times are correlated (this is
possible since $t_{ext} - t_{int}$ commutes with the total energy
$E_{total}$). Then the internal measurement of the internal energy can
also be viewed as saying that the internal observer is measuring the
energy of the {\em external} system. 

But we know that when an external
observer measures the energy of a system the internal clock gets
randomized (since the energy of the system and the clock variable are
conjugate operators in the usual sense). Now in the previous example
the internal observer can be considered ``external'' to the original
external system which now becomes the ``internal'' system. Hence the
original internal observer's measurement must randomize the original
external time. Thus eq. (\ref{DEDTMP}) must always hold.

Does this imply that eq. (\ref{DETMP}) is also universal? At first
sight one would think so. But so far we have been unable to find a
proof. (For instance it seems logically possible -although unlikely-
that one could devise a model in which the
measurement almost always takes a very short external time, and very
rarely takes a very long external time, in such a way that the average
duration of the measurement is arbitrarily short, 
but the spread in the durations
obeys eq. (\ref{DEDTMP}). Such a model would not satisfy eq. (\ref{DETMP})).
We thus leave the universal validity of eq. (\ref{DETMP}) as a
conjecture.

\section{Conclusion}

We have shown that it is possible for an internal observer to measure
the total energy of an isolated system in arbitrary short internal
time. 
This was done in a particular model.
An interesting question is
whether one can make the model more realistic, for instance by basing
it on the model of gravitational collapse discussed in the
introduction. In this respect one problem pointed out to us by
Y. Aharonov\cite{A} is that in our model the Hamiltonian is not
bounded from below. We are not sure to what extent taking a positive
Hamiltonian will modify our conclusions. Preliminary investigations
suggest that our main conclusions will remain unchanged.

\bigskip

{\bf Acknowledgment.} We thank Yakir Aharonov for enlightening
discussions. We acknowledge financial support from the European Union
through project RESQ IST-2001-37559, from EPSRC U.K. through
QIPIRC, from the 
Action de Recherche Concert\'ee de la Communaut\'e Fran{\c c}aise de
Belgique, from the IUAP program of the Belgian Federal Government
under grant V-18.

\bigskip
{\bf Figure 1: caption} The function $g(x)$ that controls when the
measuring device is coupled to the total energy.

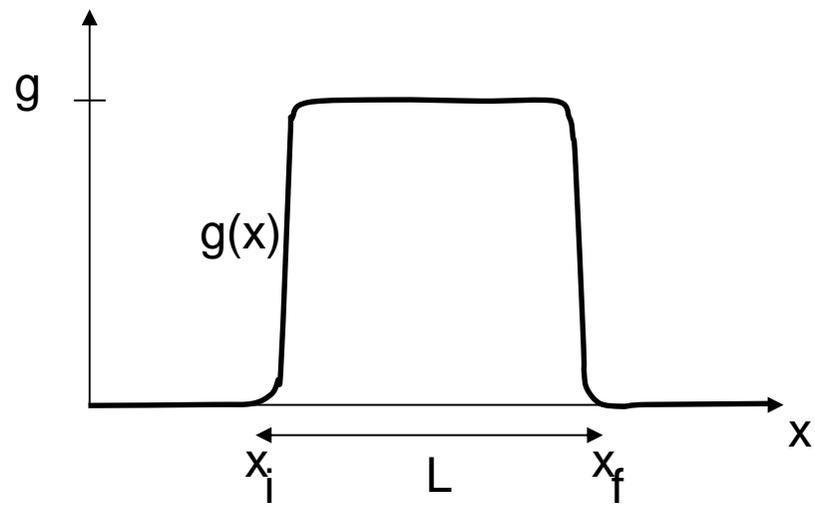
\end{document}